\documentclass{ifacconf}

\usepackage{graphicx}      
\usepackage{natbib}        
\usepackage{amssymb}
\newcommand{\be}{\begin{equation}}
\newcommand{\ee}{\end{equation}}

\begin{document}
\begin{frontmatter}

\title{Fractional Maps and Fractional Attractors. Part II:
Fractional Difference Caputo $\alpha$-Families of Maps
}

\thanks[footnoteinfo]{The author acknowledges support from the Joseph
Alexander Foundation, Yeshiva University.}

\author[yu,ci]{Edelman M.}
\address[yu]{Department of Physics, Stern College at Yeshiva University,
245 Lexington Ave, New York, NY 10016, USA, (e-mail: medelma1@yu.edu).} 
\address[ci]{Courant Institute of Mathematical Sciences,
New York University, 251 Mercer St., New York, NY 10012, USA, 
(e-mail:edelman@cims.nyu.edu).}

\begin{abstract}
In this paper we extend the notion of an
$\alpha$-family of maps to discrete systems defined 
by simple difference equations with the fractional Caputo difference operator. 
The equations considered  are equivalent to maps with falling 
factorial-law memory which is asymptotically power-law memory. 
We introduce the fractional difference Universal, Standard, and Logistic 
$\alpha$-Families of Maps
and propose to use them to study  general properties 
of discrete nonlinear systems with asymptotically power-law memory.
\end{abstract}

\end{frontmatter}

\section{Introduction}

Systems with memory are common  in biology, social sciences, physics, and 
engineering (see review \cite{MyReview}). 
The most frequently encountered
type of memory in natural and engineering systems is power-law memory.
This leads to the possibility of
describing  them by fractional differential equations,
which have power-law kernels.
Nonlinear integro-differential
fractional equations are difficult to simulate numerically - this is why
in  \cite{TZ} the authors introduced fractional maps, which
are equivalent to fractional differential equations of nonlinear
systems experiencing periodic delta function-kicks, and proposed to
use them for the investigation of general properties of 
nonlinear fractional dynamical systems. Bifurcation diagrams in the 
fractional Logistic Map related to a scheme of numerical integration 
of fractional differential equations were considered by \cite{Stan}.

An adequate description of discrete natural systems with memory    
can be obtained by using fractional difference equations 
(see \cite{MR,GZ,Aga,Atici,Anastas,DifSum,Fall,FallC}).
In \cite{DifSum,Fall,FallC} the authors
demonstrated that in some cases fractional difference equations are equivalent 
to maps (which we will call fractional
difference maps) with falling factorial-law memory, where 
falling factorial function is defined as 
\begin{equation}
t^{(\alpha)} =\frac{\Gamma(t+1)}{\Gamma(t+1-\alpha)}.
\label{FrFac}
\end{equation}
Falling factorial-law memory is asymptotically power-law memory -
\begin{equation}
\lim_{t \rightarrow
  \infty}\frac{\Gamma(t+1)}{\Gamma(t+1-\alpha)t^{\alpha}}=1,  
\ \ \ \alpha \in  \mathbb{R},
\label{GammaLimit}
\end{equation}
and we may expect that fractional difference maps have properties 
similar to the  properties of fractional maps. 

The goal of the present paper is to introduce  fractional
difference families of maps depending on memory 
and nonlinearity parameters
consistent with the previous research of
fractional maps (see
\cite{MyReview,TZ,FM1,FM2,FM3,FM4,FM5,FM6,MEDNC,MEChaos}) in order
to prepare a background for an investigation of  
general properties of systems with asymptotically power-law memory.

In the next section (Sec.~\ref{FFDFM}) we will remind the reader how 
the regular Universal, Standard, and Logistic Maps (see 
\cite{Chirikov,LL,ZasBook,May}) are generalized to obtain
fractional Caputo $\alpha$-Families of Maps ($\alpha$FM). In Sec.~\ref{FDO}
we'll present some basics on fractional difference/sum operators,
which will be used in Sec.~\ref{FDFM} to derive 
the fractional difference Caputo   Universal, Standard, and Logistic 
$\alpha$FMs. In Sec.~\ref{con} we'll present some results 
on properties of fractional difference Caputo Standard $\alpha$FM.

\section{Fractional 
$\alpha$-Families of Maps}
\label{FFDFM}


Fractional $\alpha$FM were introduced in \cite{MEDNC}, 
further investigated in \cite{MEChaos}, and reviewed in \cite{MyReview}.
The Universal $\alpha$FM was obtained by integrating the following equation:  
\begin{equation}
\frac{d^{\alpha}x}{dt^{\alpha}}+G_K(x(t- \Delta)) \sum^{\infty}_{k=-\infty} \delta \Bigl(t-(k+\varepsilon)
\Bigr)=0,   
\label{UM1D2Ddif}
\end{equation}                                                       
where $\varepsilon > \Delta > 0$,  $\alpha \in \mathbb{R}$, $\alpha>0$, 
$\varepsilon  \rightarrow 0$, with the initial conditions corresponding to
the type of fractional derivative to be used. $G_K(x)$ is a nonlinear
function which depends on the nonlinearity parameter $K$. It is called
Universal because integration of Eq.~(\ref{UM1D2Ddif}) in the case 
$\alpha =2$ and $G_K(x)=KG(x)$ produces the regular Universal Map 
(see \cite{ZasBook}).
In what follows the author considers Eq.~(\ref{UM1D2Ddif}) 
with the left-sided Caputo
fractional derivative 
(see \cite{SKM,KST,Podlubny}) 
{\setlength\arraycolsep{0.5pt}
\begin{eqnarray}
&&_0^CD^{\alpha}_t x(t)=_0I^{N-\alpha}_t \ D^N_t x(t) \nonumber \\  
&&=\frac{1}{\Gamma(N-\alpha)}  \int^{t}_0 
\frac{ D^N_{\tau}x(\tau) d \tau}{(t-\tau)^{\alpha-N+1}},  \quad ( N=\lceil \alpha \rceil),
\label{Cap}
\end{eqnarray}
}
where $N \in \mathbb{Z}$,  
$D^N_t=d^N/dt^N$, $ _0I^{\alpha}_t$ is a fractional integral,
$\Gamma()$ is the gamma function,
and the initial conditions are
\begin{equation}
(D^{k}_tx)(0+)=b_k, \ \ \ k=0,...,N-1. 
\label{UM1D2DdifIC}
\end{equation} 
There are two reasons to restrict the consideration 
in this paper to the Caputo case
(the Riemann-Liouville case 
won't be considered): 
a) as in the case of fractional differential equations, in the case of
fractional difference equations it is much easier to define 
initial conditions for  Caputo difference equations than for 
Riemann-Liouville difference equations;
b) the main goal of this work is to compare fractional 
and fractional difference maps, and the case of Caputo maps serves the
purpose.  
Comparison of the Riemann-Liouville  and Caputo Standard Maps was
considered in \cite{FM5}.  

The problem Eqs.~(\ref{UM1D2Ddif})--(\ref{UM1D2DdifIC})  
is equivalent to the Volterra integral 
equation of the second kind ($t>0$) \cite{KST}
{\setlength\arraycolsep{0.5pt}
\begin{eqnarray}
&&x(t)= \sum^{N-1}_{k=0}\frac{b_k}{k!}t^{k} \nonumber \\ 
&&\hspace{-0.4cm}-\frac{1}{\Gamma(\alpha)} \int^{t}_0 d \tau \frac{G_K(x( \tau - \Delta ))}{( t-\tau )^{1-\alpha}} \sum^{\infty}_{k=-\infty}
\delta \Bigl(\tau-(k+\varepsilon)\Bigr).
\label{VoltC}
\end{eqnarray}
}
After the introduction $x^{(s)}(t)=D^s_tx(t)$ the Caputo Universal $\alpha$FM 
can be written as (see \cite{FM4})
{\setlength\arraycolsep{0.5pt}
\begin{eqnarray}
&&x^{(s)}_{n+1}= \sum^{N-s-1}_{k=0}\frac{x^{(k+s)}_0}{k!}(n+1)^{k} \nonumber \\ 
&&-\frac{1}{\Gamma(\alpha-s)}\sum^{n}_{k=0} G_K(x_k) (n-k+1)^{\alpha-s-1},
\label{FrCMapx}
\end{eqnarray} 
}
where $s=0,1,...,N-1$ and $x^{(k+s)}_0=b_{k+s}$.

In the case $G_K(x)= K\sin(x)$ and $\alpha=2$ with $p= x^{(1)}$ 
Eq.~(\ref{FrCMapx}) produces
the well--known Standard Map (see \cite{Chirikov}),
which on a torus can be written as
\begin{equation}
p_{n+1}= p_{n} - K \sin(x_n), \ \ \ ({\rm mod} \ 2\pi ), 
\label{SMp}
\end{equation}
\begin{equation}
x_{n+1}= x_{n}+ p_{n+1}, \ \ \ ({\rm mod} \ 2\pi ).
\label{SMx}
\end{equation}
This is why the Caputo Universal $\alpha$FM Eq.~(\ref{FrCMapx}) with
\begin{equation}
G_K(x)=K \sin(x)
\label{SFM}
\end{equation} 
is called 
the Caputo Standard $\alpha$FM:
{\setlength\arraycolsep{0.5pt}
\begin{eqnarray}
&&x^{(s)}_{n+1}= \sum^{N-s-1}_{k=0}\frac{x^{(k+s)}_0}{k!}(n+1)^{k} \nonumber \\ 
&&-\frac{K}{\Gamma(\alpha-s)}\sum^{n}_{k=0} \sin(x_k) (n-k+1)^{\alpha-s-1},
\label{FrCSMMapx}
\end{eqnarray} 
}
where $s=0,1,...,N-1$.

In the case $G_K(x)= x-Kx(1-x)$ and $\alpha=1$  
Eq.~(\ref{FrCMapx}) produces
the well--known Logistic Map (see \cite{May})
\begin{equation}
x_{n+1}= Kx_{n}(1-x_{n}). 
\label{LogEq}
\end{equation}
This is why the Caputo Universal $\alpha$FM Eq.~(\ref{FrCMapx}) with
\begin{equation}
G_K(x)=G_{LK}(x)=x-Kx(1-x)
\label{LFM}
\end{equation} 
is called 
the Caputo Logistic $\alpha$FM:
{\setlength\arraycolsep{0.5pt}
\begin{eqnarray}
&&x^{(s)}_{n+1}= \sum^{N-s-1}_{k=0}\frac{x^{(k+s)}_0}{k!}(n+1)^{k} \nonumber \\ 
&&-\frac{1}{\Gamma(\alpha-s)}\sum^{n}_{k=0} \frac{x-Kx(1-x)}
{(n-k+1)^{1+s-\alpha}},
\label{FrCLMMap}
\end{eqnarray} 
}
where $s=0,1,...,N-1$.

The Caputo Standard and Logistic $\alpha$FMs were investigated in detail 
in \cite{MyReview,MEDNC,MEChaos} for the
case $\alpha \in (0,2]$  which is
important in applications.
\begin{itemize}
\item{
For $\alpha=0$ the Caputo Standard and Logistic $\alpha$FMs are 
identically zeros: $x_n=0$.
}
\item{
For $0<\alpha <1$ the Caputo Standard $\alpha$FM  is           
\begin{equation}
x_{n}=  x_0- 
\frac{K}{\Gamma(\alpha)}\sum^{n-1}_{k=0} \frac{\sin{(x_k)}}{(n-k)^{1-\alpha}},
 \   \  ({\rm mod} \ 2\pi ).
\label{FrCMapSM}
\end{equation}
and the Caputo Logistic $\alpha$FM  is
\begin{equation}
x_{n}=  x_0- 
\frac{1}{\Gamma(\alpha)}\sum^{n-1}_{k=0} \frac{x-Kx(1-x)}{(n-k)^{1-\alpha}}.
\label{FrCMapLM}
\end{equation}}
\item{
For $\alpha=1$  the 1D Standard Map 
is the Circle Map with zero driving phase
\begin{equation}
x_{n+1}= x_n - K \sin (x_n), \ \ \ \ ({\rm mod} \ 2\pi ). 
\label{SM1D} 
\end{equation}
and the  1D Logistic $\alpha$FM is the Logistic Map Eq.~(\ref{LogEq}).
}
\item{
For $1<\alpha <2$ the Caputo Standard $\alpha$FM is
{\setlength\arraycolsep{0.5pt}  
\begin{eqnarray}
&& p_{n+1} = p_n - 
 \frac{K}{\Gamma (\alpha -1 )} 
\Bigl[ \sum_{i=0}^{n-1} V^2_{\alpha}(n-i+1) \sin (x_i)  \nonumber \\
&& + \sin (x_n) \Bigr],\ \ ({\rm mod} \ 2\pi ),  \label{FSMCp}  \\ 
&& x_{n+1}=x_n+p_0-
\frac{K}{\Gamma (\alpha)} 
\sum_{i=0}^{n} V^1_{\alpha}(n-i+1) \sin (x_i), \nonumber \\ 
&& ({\rm mod} \ 2\pi ), 
\label{FSMCx}
\end{eqnarray}
}
where $V^k_{\alpha}(m)=m^{\alpha -k}-(m-1)^{\alpha -k}$
and the Caputo Logistic $\alpha$FM is
{\setlength\arraycolsep{0.5pt}
\begin{eqnarray}
&&x_{n+1}=x_0+ p(n+1)^{k} - \frac{1}{\Gamma(\alpha)}\sum^{n}_{k=0} [x_k-
\nonumber \\
&&Kx_k(1-x_k)] (n-k+1)^{\alpha-1}, 
\label{LMCx}  \\
&&p_{n+1}=p_0 - \frac{1}{\Gamma(\alpha-1)}\sum^{n}_{k=0} [x_k- \nonumber \\
&&Kx_k(1-x_k)] (n-k+1)^{\alpha-2}.
\label{LMCp}
\end{eqnarray} }}
\item{
For $\alpha=2$  the Caputo Standard Map is the regular Standard Map as in
Eqs.~(\ref{SMp})~and~(\ref{SMx}) above.
The  2D Logistic Map is 
{\setlength\arraycolsep{0.5pt}
\begin{eqnarray}
&&p_{n+1}= p_n+Kx_n(1-x_n)-x_n,
\label{LFMalp2p} \\
&&x_{n+1}= x_n + p_{n+1}.
\label{LFMalp2x}
\end{eqnarray}
}} 
\end{itemize} 

\section{Fractional Difference/Sum Operators}
\label{FDO}
  
In this paper we will adopt the definition of the fractional 
sum ($\alpha>0$)/difference ($\alpha<0$) operator
introduced in \cite{MR} as
\begin{equation}
_a\Delta^{-\alpha}_{t}f(t)=\frac{1}{\Gamma(\alpha)} \sum^{t-\alpha}_{s=a}(t-s-1)^{(\alpha-1)} f(s).
\label{MRDef}
\end{equation}
Here $f$ is defined on  $\mathbb{N}_a$ and $_a\Delta^{-\alpha}_t$ on  
$\mathbb{N}_{a+\alpha}$, where   $\mathbb{N}_t=\{t,t+1, t+2, ...\}$, and
falling factorial $t^{(\alpha)}$ is defined by Eq.~(\ref{FrFac}).
As Miller and Ross noticed, their way to introduce the discrete 
fractional sum operator based on the Green's function approach 
is not the only way to do so. In \cite{GZ} the authors defined 
the discrete fractional sum operator generalizing the $n$-fold 
summation formula in a way 
similar to the way in which 
the fractional Riemann--Liouville integral is defined in fractional 
calculus by extending the Cauchy $n$-fold integral formula
to the real variables. They mentioned the following theorem but didn't 
present a proof. 
\begin{thm}
For $\forall n \in \mathbb{N}$   
{\setlength\arraycolsep{0.5pt}   
\begin{eqnarray} 
&&_a\Delta^{-n}_{t}f(t)=\frac{1}{(n-1)!} \sum^{t-n}_{s=a}(t-s-1)^{(n-1)}
f(s)\nonumber \\
&&=\sum^{t-n}_{s^0=a} \sum^{s^0}_{s^1=a}...
\sum^{s^{n-2}}_{s^{n-1}=a}f(s^{n-1}),
\label{MRInt}
\end{eqnarray}
} 
where $s^{i}$, $i=0,1,...n-1$ are the summation variables. 
\end{thm}
\begin{pf}
Indeed,
this formula is obviously true for $n=1$. Let's assume that 
Eq.~(\ref{MRInt}) is true for $n-1$:
{\setlength\arraycolsep{0.5pt}   
\begin{eqnarray} 
&&_a\Delta^{-(n-1)}_{t}f(t)=\sum^{t-(n-1)}_{s=a}C(t-s-1,n-2) f(s)
\nonumber \\
&&= \sum^{t-(n-1)}_{s^1=a} \sum^{s^1}_{s^2=a}...
\sum^{s^{n-2}}_{s^{n-1}=a}f(s^{n-1}),
\label{IntSum1}
\end{eqnarray}
} 
where $C(i,j)$ is the number of $j$-combinations from a given set of $i$
elements. Then, for $t=s^0+n-1$ Eq.~(\ref{IntSum1})  gives
{\setlength\arraycolsep{0.5pt}   
\begin{eqnarray} 
&&\sum^{s^0}_{s^1=a}C(s^0-s^1+n-2,n-2) f(s^1)
\nonumber \\
&&= \sum^{s^0}_{s^1=a} \sum^{s^1}_{s^2=a}...
\sum^{s^{n-2}}_{s^{n-1}=a}f(s^{n-1}).
\label{IntSum2}
\end{eqnarray}
} 
Now Eq.~(\ref{MRInt}) can be obtained from 
{\setlength\arraycolsep{0.5pt}   
\begin{eqnarray} 
&&\sum^{t-n}_{s^0=a} \sum^{s^0}_{s^1=a}...
\sum^{s^{n-2}}_{s^{n-1}=a}f(s^{n-1})
\nonumber \\
&&= \sum^{t-n}_{s^0=a}\sum^{s^0}_{s^1=a}C(s^0-s^1+n-2,n-2) f(s^1)
\nonumber \\
&&=\sum^{t-n}_{s^1=a}f(s^1) \sum^{t-n}_{s^0=s^1}C(s^0-s^1+n-2,n-2)
\nonumber \\
&&=\sum^{t-n}_{s^1=a}C(t-s^1-1,n-1) f(s^1)=_a\Delta^{-n}_{t}f(t).
\label{IntSum3}
\end{eqnarray}
} 
Here we used the identity
\begin{equation}
\sum^{t-n}_{s^0=s^1}C(s^0-s^1+n-2,n-2)
=C(t-s^1-1,n-1),
\label{CombIdent}
\end{equation}
which is true for $t=n+s^1$ and can be proven by induction for any
$t$ 
{\setlength\arraycolsep{0.5pt}   
\begin{eqnarray} 
&&\sum^{t+1-n}_{s^0=s^1}C(s^0-s^1+n-2,n-2)= C(t-s^1-1,n-2)
\nonumber \\
&&+C(t-s^1-1,n-1)=C(t-s^1,n-1).
\label{CombIdentN}
\end{eqnarray}
} 
This ends the proof.
\end{pf}

As we see, two different approaches are consistent with the  definition
of the fractioanal sum operator given by Miller and Ross
(see also \cite{Atici}).
For $\alpha >0$ and   $m-1<\alpha < m$ 
\cite{Anastas} defined the fractional (left) Caputo-like difference operator as
{\setlength\arraycolsep{0.5pt}   
\begin{eqnarray} 
&&_a^C\Delta^{\alpha}_t x(t) =  _a\Delta^{-(m-\alpha)}_{t}\Delta^{m} x(t)\nonumber \\
&& =\frac{1}{\Gamma(m-\alpha)} \sum^{t-(m-\alpha)}_{s=a}(t-s-1)^{(m-\alpha-1)} 
\Delta^m x(s),
\label{FDC}
\end{eqnarray}
}
where $\Delta^{m}$ is the $m$-th power of the forward difference operator
defined as $\Delta x(t)=x(t+1)-x(t)$. 
The  proof (see \cite{MR} p.146) that $_0\Delta^{\lambda}_t$ in the limit 
$\lambda \rightarrow 0$  approaches the identity operator can be easily
extended to  the $_a\Delta^{\lambda}_t$ operator. In this case the 
definition Eq.~(\ref{FDC}) can be extended to all real  $\alpha \ge 0$
with $_a^C\Delta^{m}_t x(t) = \Delta^m x(t)$ for $m \in \mathbb{N}_0$.
Then, the Anastassiou's fractional Taylor difference formula \cite{Anastas}
{\setlength\arraycolsep{0.5pt}   
\begin{eqnarray} 
&&x(t) =   \sum^{m-1}_{k=0}\frac{(t-a)^{(k)}}{k!}\Delta^{k}x(a) \nonumber \\
&& +\frac{1}{\Gamma(\alpha)} \sum^{t-\alpha}_{s=a+m-\alpha}(t-s-1)^{(\alpha-1)} 
 {_a^C\Delta^{\alpha}_t} x(t),
\label{FTaylor}
\end{eqnarray}
}
where $x$ is defined on $ \mathbb{N}_{a}$,  $m=\lceil \alpha \rceil$, and 
$a \in \mathbb{N}_{0}$ for  $\forall t \in \mathbb{N}_{a+m}$ is valid for 
any real $ \alpha >0$ and for integer  $ \alpha = m$ is identical to
the integer discrete Taylor's formula (see p.28 in \cite{Aga})
{\setlength\arraycolsep{0.5pt}   
\begin{eqnarray} 
&&x(t) =   \sum^{m-1}_{k=0}\frac{(t-a)^{(k)}}{k!}\Delta^{k}x(a) \nonumber \\
&& +\frac{1}{(m-1)!} \sum^{t-m}_{s=a}(t-s-1)^{(m-1)} 
 \Delta^{m} x(t).
\label{DTaylor}
\end{eqnarray}
}

As it was noticed in \cite{Fall} and \cite{FallC}, Lemma 2.4 from \cite{DifSum}
on the equivalency of the fractional Caputo-like difference 
and sum equations can be extended to all real $\alpha>0$ and 
formulated as follows:
\begin{thm}
The Caputo-like difference equation 
\begin{equation}
_a^C\Delta^{\alpha}_t x(t) = f(t+\alpha-1,x(t+\alpha-1)
\label{LemmaDif}
\end{equation}
with the initial conditions 
 \begin{equation}
\Delta^{k} x(a) = c_k, \ \ \ k=0, 1, ..., m-1, \ \ \ 
m=\lceil \alpha \rceil
\label{LemmaDifIC}
\end{equation}
is equivalent to the fractional sum equation
{\setlength\arraycolsep{0.5pt}   
\begin{eqnarray} \label{LemmaSum}
&&x(t) =   \sum^{m-1}_{k=0}\frac{(t-a)^{(k)}}{k!}\Delta^{k}x(a) 
+\frac{1}{\Gamma(\alpha)}  \\
&& \times \sum^{t-\alpha}_{s=a+m-\alpha}(t-s-1)^{(\alpha-1)} 
f(s+\alpha-1,x(s+\alpha-1)), \nonumber
\end{eqnarray}
}
where $t\in \mathbb{N}_{a+m}$.
\label{Lemma}
\end{thm}
Here we should notice that the authors of \cite{Fall} and \cite{FallC} 
didn't consider the Caputo difference operator with integer $\alpha$.  
As a result, Theorem~\ref{Lemma} is not valid for integer values 
of $\alpha$ with their definition $m=[\alpha]+1$.

This theorem in the limiting sense  can be extended to all real $\alpha \ge 0$.
Indeed, taking into account that ${\displaystyle\lim_{\alpha \to 0}}
{_a^C\Delta^{\alpha}_t x(t)}=x(t)$,  Eq.~(\ref{LemmaDif}) for $\alpha=0$ 
turns into 
\begin{equation}
x(t) = f(t-1,x(t-1).
\label{LemmaDif0}
\end{equation} 
For  $\alpha=0$ the first sum on the right in  Eq.~(\ref{LemmaSum}) 
disappears and in the second sum the only remaining term with $s=t-\alpha$
in the limit $\alpha \rightarrow 0$ turns into  $f(t-1,x(t-1))$.

\section{Fractional Difference 
$\alpha$-Families of Maps}
\label{FDFM}
  
In the following we assume that $f$ is a  
nonlinear function $f(t,x(t))=-G_K(x(t))$ with the nonlinearity 
parameter $K$ and adopt the Miller and Ross proposition to let $a=0$.
Now, with $x_n=x(n)$, Theorem~\ref{Lemma} can be formulated as
\begin{thm}
 For $\alpha \in \mathbb{R}$, $\alpha \ge 0$ the Caputo-like 
difference equation 
\begin{equation}
_0^C\Delta^{\alpha}_t x(t) = -G_K(x(t+\alpha-1)),
\label{LemmaDif_n}
\end{equation}
where $t\in \mathbb{N}_{m}$, with the initial conditions 
 \begin{equation}
\Delta^{k} x(0) = c_k, \ \ \ k=0, 1, ..., m-1, \ \ \ 
m=\lceil \alpha \rceil
\label{LemmaDifICn}
\end{equation}
is equivalent to the map with falling factorial-law memory
{\setlength\arraycolsep{0.5pt}   
\begin{eqnarray} 
&&x_{n+1} =   \sum^{m-1}_{k=0}\frac{\Delta^{k}x(0)}{k!}(n+1)^{(k)} 
\nonumber \\
&&-\frac{1}{\Gamma(\alpha)}  
\sum^{n+1-m}_{s=0}(n-s-m+\alpha)^{(\alpha-1)} 
G_K(x_{s+m-1}), 
\label{FalFacMap}
\end{eqnarray}
}
where $x_k=x(k)$ which we will call 
the  fractional difference Caputo  Universal  $\alpha$-Family of Maps.
\end{thm}
The  fractional difference Caputo  Universal  $\alpha$FM
is similar to the general form of the Caputo Universal
$\alpha$FM Eq.~(\ref{FrCMapx}). Both of them can be written as
{\setlength\arraycolsep{0.5pt}   
\begin{eqnarray} 
&&x_{n} =   x_0+\sum^{m-1}_{k=1}\frac{p^k(0)}{k!}n^{(k)} 
\nonumber \\
&&-\frac{1}{\Gamma(\alpha)}\sum^{n-1}_{k=M} W_{\alpha}(n-k) G_K{(x_k)},
\label{UnivFrFrdif}
\end{eqnarray}
}
where $p^k(0)$ are the initial value of momenta defined as
$p^s=D_t^sx(t)$ for fractional maps and as $p^s(t)=\Delta^sx(t)$
for fractional difference maps; $M=0$ for fractional maps and
$M=m-1$ for fractional difference maps;
$n^{(s)}=n^s$ for fractional maps and  $n^{(s)}=\Gamma(n+1)/\Gamma(n+1-s)$
for the fractional difference maps.
$W_{\alpha}(s) = s^{\alpha-1}$ for fractional maps and
 $W_{\alpha}(s) ={\Gamma(s+\alpha-1)}/{\Gamma(s)}$ for  fractional
 difference maps. Asymptotically, both expressions for $W_{\alpha}(s)$ 
coincide because of Eq.~(\ref{GammaLimit}).

\subsection{Fractional Difference Universal $\alpha$FM}
\label{Alp2}

Let's consider the case $\alpha=2$. Then the difference 
Eq.~(\ref{LemmaDif}) produces
\begin{equation}
\Delta^{2} x_n = -G_K(x_{n+1})
\label{UMDif}
\end{equation}
and the equivalent sum equation is
\begin{equation}
x_{n+1} =x_0+\Delta x_0 (n+1)
-\sum^{n-1}_{s=0}(n-s)G_K(x_{s+1}).
\label{UMSum} 
\end{equation}
After the introduction $p_n=\Delta x_{n-1}$ with the assumption
$G_K(x)=KG(x)$ 
the map equations indeed can be written as the well--known 2D Universal Map    
\begin{equation}
p_{n+1}= p_{n} - K G(x_n),  
\label{UMp}
\end{equation}
\begin{equation}
x_{n+1}= x_{n}+ p_{n+1}, 
\label{UMx}
\end{equation}
which for $G(x)=\sin(x)$ produces the Standard Map 
Eqs.~(\ref{SMp})~and~(\ref{SMx}).
In the rest of this paper we'll call  Eq.~(\ref{FalFacMap}) 
with $G_K(x)=K\sin(x)$  the  
fractional difference  Caputo 
Standard $\alpha$-Family of Maps.

In the case $\alpha=1$ the fractional difference Caputo  Universal
$\alpha$FM is
\begin{equation}
x_{n+1}= x_{n}- G_K(x_n), 
\label{UM1D}
\end{equation}
which  produces the Logistic Map if   $G_K(x)=x-Kx(1-x)$.
In the rest of this paper we'll call  Eq.~(\ref{FalFacMap}) 
with $G_K(x)=x-Kx(1-x)$  the  
fractional difference  Caputo 
Logistic $\alpha$-Family of Maps.

\subsection{$\alpha=0$  
Difference Caputo  Standard
and Logistic $\alpha$FMs   }
\begin{itemize}
\item{
In the case $\alpha=0$ the 0D Standard Map turns into the Sine Map (see,
e.g., \cite{Sin})
\begin{equation}
x_{n+1} = -K\sin(x_n), \ \ \ ({\rm mod} \ 2\pi ).
\label{SM0D}
\end{equation} 
}
\item{
The 0D Logistic Map is
\begin{equation}
x_{n+1} =-x_n+Kx_n(1-x_n).
\label{LM0D}
\end{equation} 
}
\end{itemize}

\subsection{$0<\alpha<1$ Fractional Difference  Caputo 
Standard and Logistic $\alpha$FMs}
\label{AlpLT1}
\begin{itemize}
\item{
For $0<\alpha<1$ the fractional difference Standard Map
is 
{\setlength\arraycolsep{0.5pt}   
\begin{eqnarray} 
&&x_{n+1} =  x_0  
\label{SMlt1} \\
&& -\frac{K}{\Gamma(\alpha)}
\sum^{n}_{s=0}\frac{\Gamma(n-s+\alpha)}{\Gamma(n-s+1)}\sin (x_s) 
, \ \ \ ({\rm mod} \ 2\pi ),
\nonumber
\end{eqnarray}
}
which after the $\pi$-shift of the independent variable 
$x\rightarrow x+\pi$ coincides with the ``fractional sine map''
proposed in \cite{Fall}.
}
\item{
The fractional difference Logistic Map can be writen as
{\setlength\arraycolsep{0.5pt}   
\begin{eqnarray} 
&&x_{n+1} =  x_0  
\label{LMlt1} \\
&& -\frac{1}{\Gamma(\alpha)}
\sum^{n}_{s=0}\frac{\Gamma(n-s+\alpha)}{\Gamma(n-s+1)}[x_s-Kx_s(1-x_s)]. 
\nonumber
\end{eqnarray}
}
The fractional Logistic Map introduced in \cite{FallC} 
does not converge to the Logistic map in the case $\alpha=1$. 
}
\end{itemize}

\subsection{$\alpha=1$ Difference  Caputo 
Standard  and Logistic $\alpha$FMs 
}
\label{Alp1}
\begin{itemize}
\item{The $\alpha=1$ difference  Caputo 
Standard   $\alpha$FM is identical to the Circle Map with zero driven phase
Eq.~(\ref{SM1D}). The map considered in \cite{Fall} 
\begin{equation}
x_{n+1}= x_n + K \sin (x_n), \ \ \ \ ({\rm mod} \ 2\pi )
\label{SM1DNeg} 
\end{equation}
is obtained from this map by the substitution $x \rightarrow x+\pi$.
}
\item{The $\alpha=1$ Difference  Caputo Logistic $\alpha$FM is 
the regular Logistic Map.}
\end{itemize}

\subsection{$1<\alpha<2$ Fractional Difference Caputo 
Standard and Logistic $\alpha$FMs  }
\label{Alp1to2}
\begin{itemize}
\item{
For $1<\alpha<2$ the fractional difference Standard Map
is 
{\setlength\arraycolsep{0.5pt}   
\begin{eqnarray} 
&&x_{n+1} =  x_0 +\Delta x_0 (n+1) -\frac{K}{\Gamma(\alpha)}
\label{SMgt1} \\
&& 
\times \sum^{n-1}_{s=0}\frac{\Gamma(n-s+\alpha-1)}{\Gamma(n-s)}\sin (x_{s+1}) 
, \ \ ({\rm mod} \ 2\pi ).
\nonumber
\end{eqnarray}
}
}
\item{
The $1<\alpha<2$ fractional difference Logistic Map is
{\setlength\arraycolsep{0.5pt}   
\begin{eqnarray} 
&&x_{n+1} =  x_0 +\Delta x_0 (n+1) -\frac{1}{\Gamma(\alpha)}
 \label{LMgt1} \\
&& 
\times
\sum^{n-1}_{s=0}\frac{\Gamma(n-s+\alpha-1)}{\Gamma(n-s)}[x_{s+1}-Kx_{s+1}(1-
x_{s+1})].
\nonumber
\end{eqnarray}
}
}
\end{itemize}
Let's introduce  $p_n=\Delta x_{n-1}$; then these maps can be written 
as 2D maps with memory:
\begin{itemize}
\item{
The fractional difference Standard Map is
{\setlength\arraycolsep{0.5pt}   
\begin{eqnarray} 
&&p_{n} =  p_1 -\frac{K}{\Gamma(\alpha-1)}
\label{SMgt1p} \\
&&
\times \sum^{n}_{s=2}\frac{\Gamma(n-s+\alpha-1)}
{\Gamma(n-s+1)}\sin (x_{s-1}) 
, \ \ ({\rm mod} \ 2\pi ),
\nonumber  \\
&& x_n=x_{n-1}+p_n, \ \ ({\rm mod} \ 2\pi ),  \ \ n \ge 1,
\label{SMgt1x}
\end{eqnarray}
}
which in the case $x_0=0$ is identical to the 
"fractional standard map" introduced in \cite{Fall} (Eq.~(18) with 
$\nu=\alpha-1$ there). 
}
\item{
The fractional difference Logistic Map is 
{\setlength\arraycolsep{0.5pt}   
\begin{eqnarray} 
&&p_{n} =  p_1 -\frac{K}{\Gamma(\alpha-1)}
\label{LMgt1p}  \\
&&\times \sum^{n}_{s=2}\frac{\Gamma(n-s+\alpha-1)}
{\Gamma(n-s+1)}[x_{s-1}-Kx_{s-1}(1-x_{s-1})], 
\nonumber \\
&& x_n=x_{n-1}+p_n,  \ \ n \ge 1.
\label{LMgt1x} 
\end{eqnarray}
}
}
\end{itemize}

\subsection{$\alpha=2$  Difference  Caputo 
Standard and Logistic $\alpha$FMs}
\label{SMAlp2}
\begin{itemize}
\item{The $\alpha=2$  difference  Caputo 
Standard $\alpha$FM is  
the regular Standard Map Eqs.~(\ref{SMp})~and~(\ref{SMx}).
}
\item{
From Eqs.~(\ref{UMp})~and~(\ref{UMx}) the 2D difference  Caputo 
Logistic $\alpha$FM is
\begin{equation}
p_{n+1}= p_{n} - x_n+Kx_n(1-x_n),  
\label{LMp}
\end{equation}
\begin{equation}
x_{n+1}= x_{n}+ p_{n+1}, 
\label{LMx}
\end{equation}
which is identical to the 2D Logistic Map 
Eqs.~(\ref{LFMalp2p}) and (\ref{LFMalp2x}).
}
\end{itemize} 

\section{Conclusion}
\label{con}

As we saw in Sec.~\ref{FDO}, the fractional difference operator is a natural 
extension of the difference operator. The simplest fractional difference
equations (of the Eq.~(\ref{LemmaDif}) type), where the fractional
difference on the left side is
equal to a nonlinear function on the right side, are equivalent to 
maps with falling factorial-law (asymptotically power-law) memory 
Eq.~(\ref{LemmaSum}). Systems with power-law memory play an important role
in nature (see \cite{MyReview}) and investigation of their general properties is
important for understanding behavior of  natural systems.
\begin{figure}[!t]
\centering
\includegraphics[width=3in]{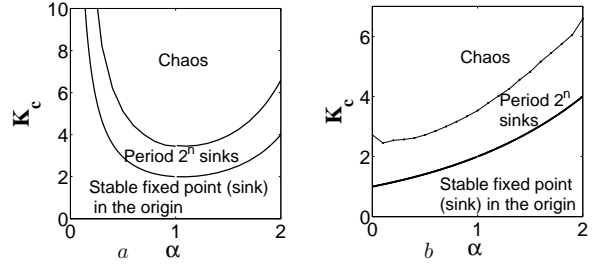}
\caption{$\alpha-K$ (bifurcation) diagrams  for the Caputo ($a$)
 and  Fractional  Difference Caputo ($b$) Standard $\alpha$FMs.
}
\label{F1}
\end{figure}
\begin{figure}[!t]
\centering
\includegraphics[width=3in]{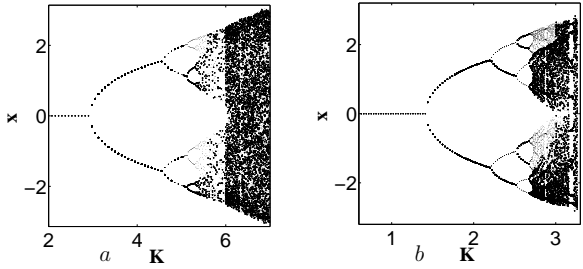}
\caption{  Bifurcation diagrams  for the Caputo ($a$)
 and  Fractional  Difference Caputo ($b$) Standard $\alpha$FMs
with  $\alpha=0.5$ obtained after 5000 iterations with the initial
condition $x_0=0.1$ (regular points) and $x_0=-0.1$ (bold points).
}
\label{F2}
\end{figure}
\begin{figure}[!t]
\centering
\includegraphics[width=3in]{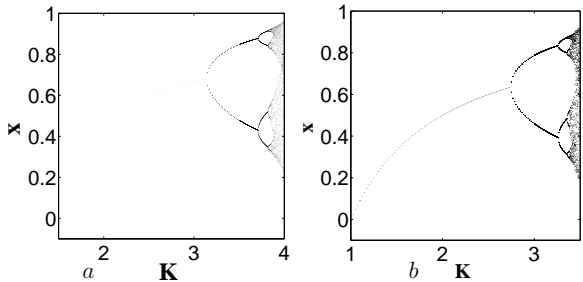}
\caption{  Bifurcation diagrams  for the Caputo ($a$)
 and  Fractional  Difference Caputo ($b$) Logistic $\alpha$FMs
with  $\alpha=0.8$ obtained after 1000 iterations with the initial
condition $x_0=0.01$.
}
\label{F3N}
\end{figure}
\begin{figure}[!t]
\centering
\includegraphics[width=3in]{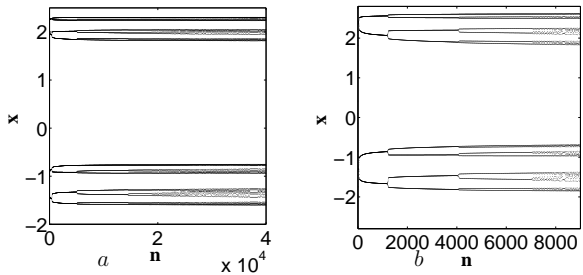}
\caption{Cascade of bifurcations type trajectories for the Caputo ($a$)
 and  Fractional  Difference Caputo ($b$) Standard $\alpha$FMs
with  $\alpha=0.1$ and $x_0=0.1$. $K=26.65$ in   ($a$) and $K=2.41$ in ($b$). 
}
\label{F3}
\end{figure}
\begin{figure}[!t]
\centering
\includegraphics[width=3in]{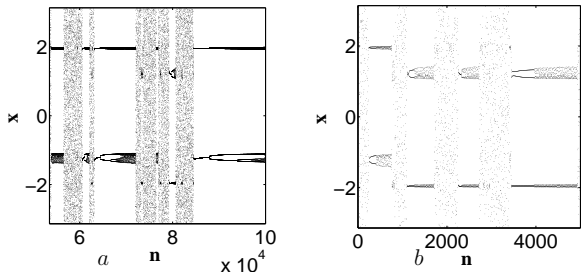}
\caption{Intermittent cascade of bifurcations type trajectories 
for the Caputo ($a$) and  Fractional  Difference Caputo ($b$) 
Standard $\alpha$FMs. In ($a$) $\alpha=1.65$,  $K=4.5$, $x_0=0.$, and
$p_0=0.3$. In ($b$) $\alpha=1.5$,  $K=4.82$, $x_0=0.$, and
$p_0=0.01$.
}
\label{F4}
\end{figure}

Properties of the fractional difference Caputo Standard $\alpha$FM were 
investigated in detail 
in \cite{MEPhysLetA} (see also Sec.~3 in \cite{MEArchive}.
Qualitatively, properties of the fractional difference and fractional maps 
(maps with falling 
factorial- and power-law memory) are similar. 
The similarity reveals itself in the dependence of systems' properties on
the memory ($\alpha$)
and nonlinearity ($K$) parameters (bifurcation
diagrams, see Figs.~\ref{F1},~\ref{F2},~and~\ref{F3N}), 
power-law convergence to attractors, 
non-uniqueness of solutions 
(intersection of trajectories and overlapping of attractors),
and cascade of bifurcations and intermittent 
cascade of bifurcations type behaviors 
(see Figs.~\ref{F3}~and~\ref{F4}).

The differences of the properties of the falling factorial-law memory maps
from the power-law memory maps are the results of the differences in
weights of the recent (with $(n-j)/n <<1$) 
values of the maps' variables at the time instants $t_j$ 
in the definition of the present values at time $t_n$ 
and are significant when $\alpha \in (0,1)$ 
(especially when $\alpha \rightarrow +0$), 
see Fig.~\ref{F1}.
 
\begin{ack}
The author acknowledges support from the Joseph Alexander Foundation,
Yeshiva University.  
The author expresses his gratitude to E. Hameiri, H. Weitzner,
and  G. Ben Arous
for the opportunity to complete this work at the Courant Institute     
and to V. Donnelly for technical help. 

\end{ack}


\bibliography{FallMay20ifac}             

\begin{thebibliography}{29}
\providecommand{\natexlab}[1]{#1}
\providecommand{\url}[1]{\texttt{#1}}
\providecommand{\urlprefix}{URL }
\expandafter\ifx\csname urlstyle\endcsname\relax
  \providecommand{\doi}[1]{doi:\discretionary{}{}{}#1}\else
  \providecommand{\doi}{doi:\discretionary{}{}{}\begingroup
  \urlstyle{rm}\Url}\fi

\bibitem[{Agarwal(2000)}]{Aga}
Agarwal, R. (2000).
\newblock \emph{Difference equations and inequalities}.
\newblock Marcel Dekker, New York.

\bibitem[{Anastassiou(2009)}]{Anastas}
Anastassiou, G. (2009).
\newblock Discrete fractional calculus and inequalities.
\newblock \emph{http://arxiv.org/abs/0911.3370}.

\bibitem[{Atici and Eloe(2009)}]{Atici}
Atici, F. and Eloe, P. (2009).
\newblock Initial value problems in discrete fractional calculus.
\newblock \emph{Proc.Am.Math.Soc.}, 137, 981--989.

\bibitem[{Chen et~al.(2011)Chen, Luo, and Zhou}]{DifSum}
Chen, F., Luo, X., and Zhou, Y. (2011).
\newblock Existence results for nonlinear fractional difference equation.
\newblock \emph{Adv.Differ.Eq.}, 2011, 713201.

\bibitem[{Chirikov(1979.)}]{Chirikov}
Chirikov, B. (1979.).
\newblock A universal instability of many dimensional oscillator systems.
\newblock \emph{Phys. Rep.}, 52, 263--379.

\bibitem[{Edelman(2011)}]{FM5}
Edelman, M. (2011).
\newblock Fractional standard map: Riemann-liouville vs. caputo.
\newblock \emph{Commun. Nonlin. Sci. Numer. Simul.}, 16, 4573--4580.

\bibitem[{Edelman(2013{\natexlab{a}})}]{MEDNC}
Edelman, M. (2013{\natexlab{a}}).
\newblock Fractional maps and fractional attractors. part i: $\alpha$-families
  of maps.
\newblock \emph{Discontinuity, Nonlinearity, and Complexity}, 1, 305--324.

\bibitem[{Edelman(2013{\natexlab{b}})}]{MEChaos}
Edelman, M. (2013{\natexlab{b}}).
\newblock Universal fractional map and cascade of bifurcations type attractors.
\newblock \emph{Chaos}, 23, 033127.

\bibitem[{Edelman(2014{\natexlab{a}})}]{MEPhysLetA}
Edelman, M. (2014{\natexlab{a}}).
\newblock Caputo standard $\alpha$-family of maps: fractional difference vs.
  fractional.
\newblock \emph{Phys. Lett. A (submitted)}.

\bibitem[{Edelman(2014{\natexlab{b}})}]{MEArchive}
Edelman, M. (2014{\natexlab{b}}).
\newblock Fractional maps and fractional attractors. part ii: fractional
  difference $\alpha$-families of maps.
\newblock \emph{arXiv:1404.4906v2}.

\bibitem[{Edelman(2014{\natexlab{c}})}]{MyReview}
Edelman, M. (2014{\natexlab{c}}).
\newblock Fractional maps as maps with power-law memory.
\newblock In A.~Afraimovich, A.~Luo, and X.~Fu (eds.), \emph{Nonlinear dynamics
  and complexity}, 79--120. Springer, New York.

\bibitem[{Edelman and Taieb(2013)}]{FM6}
Edelman, M. and Taieb, L. (2013).
\newblock New types of solutions of non-linear fractional differential
  equations.
\newblock In A.~Almeida, L.~Castro, and F.O. Speck (eds.), \emph{Advances in
  Harmonic Analysis and Operator Theory; Series: Operator Theory: Advances and
  Applications}, volume 229, 139--155. Springer, Basel.

\bibitem[{Edelman and Tarasov(2009)}]{FM1}
Edelman, M. and Tarasov, V.E. (2009).
\newblock Fractional standard map.
\newblock \emph{Phys. Lett. A}, 374, 279--285.

\bibitem[{Gray and Zhang(1988)}]{GZ}
Gray, H. and Zhang, N.F. (1988).
\newblock On a new definition of the fractional difference.
\newblock \emph{Math. Comput.}, 50, 513--529.

\bibitem[{Kilbas et~al.(2006)Kilbas, Srivastava, and Trujillo}]{KST}
Kilbas, A., Srivastava, H., and Trujillo, J. (2006).
\newblock \emph{Theory and application of fractional differential equations}.
\newblock Elsevier, Amsterdam.

\bibitem[{Lalescu(2010)}]{Sin}
Lalescu, C. (2010).
\newblock Patterns in the sine map bifurcation diagram.
\newblock \emph{arXiv:1011.6552}.

\bibitem[{Lichtenberg and Lieberman(1992)}]{LL}
Lichtenberg, A. and Lieberman, M. (1992).
\newblock \emph{Regular and chaotic dynamics}.
\newblock Springer, Berlin.

\bibitem[{May(1976)}]{May}
May, R. (1976).
\newblock Simple mathematical models with very complicated dynamics.
\newblock \emph{Nature}, 261, 459--467.

\bibitem[{Miller and Ross(1989)}]{MR}
Miller, K. and Ross, B. (1989).
\newblock Fractional difference calculus.
\newblock In H.~Srivastava and S.~Owa (eds.), \emph{Univalent functions,
  fractional calculus, and their applications}, 139--151. Ellis Howard,
  Chichester, 1st edition.

\bibitem[{Podlubny(1999)}]{Podlubny}
Podlubny, I. (1999).
\newblock \emph{Fractional Differential Equations}.
\newblock Academic Press, San Diego.

\bibitem[{Samko et~al.(1993)Samko, Kilbas, and Marichev}]{SKM}
Samko, S., Kilbas, A., and Marichev, O. (1993).
\newblock \emph{fractional integrals and derivatives theory and applications}.
\newblock Gordon and Breach, New York.

\bibitem[{Stanislavsky(2006)}]{Stan}
Stanislavsky, A. (2006).
\newblock Long-term memory contribution as applied to the motion of discrete
  dynamical system.
\newblock \emph{Chaos}, 16, 043105.

\bibitem[{Tarasov(2009{\natexlab{a}})}]{FM2}
Tarasov, V. (2009{\natexlab{a}}).
\newblock Differential equations with fractional derivative and universal map
  with memory.
\newblock \emph{J. Phys. A}, 42, 465102.

\bibitem[{Tarasov(2009{\natexlab{b}})}]{FM3}
Tarasov, V. (2009{\natexlab{b}}).
\newblock Discrete map with memory from fractional differential equation of
  arbitrary positive order.
\newblock \emph{J. Math. Phys.}, 50, 122703.

\bibitem[{Tarasov(2011)}]{FM4}
Tarasov, V. (2011).
\newblock \emph{Fractional dynamics: application of fractional calculus to
  dynamics of particles, fields, and media}.
\newblock Springer, New York.

\bibitem[{Tarasov and Zaslavsky(2008)}]{TZ}
Tarasov, V. and Zaslavsky, G. (2008).
\newblock Fractional equations of kicked systems and discrete maps.
\newblock \emph{J. Phys. A}, 41, 435101.

\bibitem[{Wu and Baleanu(2014)}]{FallC}
Wu, G.C. and Baleanu, D. (2014).
\newblock Discrete fractional logistic map and its chaos.
\newblock \emph{Nonlin. Dyn.}, 75, 283--287.

\bibitem[{Wu et~al.(2014)Wu, Baleanu, and Zeng}]{Fall}
Wu, G.C., Baleanu, D., and Zeng, S.D. (2014).
\newblock Discrete chaos in fractional sine and standard maps.
\newblock \emph{Phys. Lett. A}, 378, 484--487.

\bibitem[{Zaslavsky(2008)}]{ZasBook}
Zaslavsky, G. (2008).
\newblock \emph{Hamiltonian chaos and fractional dynamics}.
\newblock Oxford University Press, Oxford.

\end{thebibliography}

\end{document}